\documentclass[prd,twocolumn,floatfix,preprintnumbers,letterpaper]{revtex4}

\usepackage{graphicx}
\input{epsf}
\usepackage{epsf}
\usepackage{graphicx,epsfig}
\usepackage{bm}
\usepackage{latexsym,amssymb,amsmath,float}

\newcommand {\ga} {\ {\raise-.5ex\hbox{$\buildrel>\over\sim$}}\ }
\newcommand {\la} {\ {\raise-.5ex\hbox{$\buildrel<\over\sim$}}\ } 

\begin{document}

\title{Evolution of Oscillating Scalar Fields as Dark Energy}
\author{Sourish Dutta and Robert J. Scherrer}
\affiliation{Department of Physics and Astronomy, Vanderbilt University,
Nashville, TN  ~~37235}

\begin{abstract}
Oscillating scalar fields, with an oscillation frequency
much greater than the expansion rate, have been proposed as models for dark energy.
We examine these models, with particular emphasis on the evolution
of the ratio of the oscillation frequency to the expansion rate.
We show that this ratio always increases with time if the dark energy
density declines less rapidly than the background energy density.  This
allows us to classify oscillating dark energy models in terms of
the epoch at which the oscillation frequency exceeds the expansion rate,
which is effectively the time at which rapid oscillations begin.
There are three basic types of behavior:
early oscillation models, in which oscillations begin during
the matter-dominated era, late oscillation models, in which oscillations
begin after scalar-field domination, and non-oscillating models.
We examine a representative set of models (those with power-law potentials)
and determine the parameter range giving acceptable agreement with
the supernova observations.  We show that a subset of all three
classes of models can be consistent with the observational data.
\end{abstract}

\maketitle

\section{Introduction}

It appears that
approximately 70\% of the energy density in the
universe is in the form of an exotic, negative-pressure component,
dubbed dark energy \cite{Knop,Riess1,Wood-Vasey,Davis}.
(See Ref. \cite{Copeland}
for a recent review).
The most likely possibility for this dark energy is a cosmological constant.
Another possibility,
dubbed quintessence, is a model in which the dark energy
arises from a scalar
field \cite{ratra,tw,caldwelletal,liddle,zlatev}.

Although it is usually assumed that the evolution of the scalar
field is monotonic, a more exotic possibility is that the scalar
field oscillates.  Such oscillating scalar fields were first
investigated by Turner \cite{turner}, and later reexamined in the context
of models for dark energy \cite{liddle,sahni,hsu,masso,gu}.  (See
also Ref. \cite{KHS} for a phantom version of these models,
and Ref. \cite{DM} for applications to inflation).
Here we examine a key aspect of
such models:  the interplay between the oscillation frequency
and the Hubble expansion rate.  The discussions
in Refs. \cite{turner,sahni,hsu,masso,gu} assumed that the oscillation
frequency is large compared to the expansion rate.
However, if the ratio between these two
quantities crosses unity, it is possible to generate interesting
new behaviors for the dark energy.  We explore this possibility here.
(Note that the models considered here, in which the scalar field oscillates,
are distinct from models in which the equation of state oscillates
\cite{dodosc,barosc1,barosc2,fengosc,linderosc}.  In the
latter case, one is generally interested in oscillation frequencies
which are always
on the order of the Hubble parameter.  In the models
considered here, oscillations with frequencies on the order of the Hubble
parameter are a transient, albeit important, phenomenon).

In the next section, we review the evolution of oscillating
scalar fields in an expanding background, and we derive some
interesting general results on the evolution of the ratio
of scalar field oscillation frequency to the expansion rate.
In Sec. III, we compare numerical simulations of a representative
set of oscillating models (due to power-law potentials) with
the observations.  Our conclusions are summarized in Sec. IV.
Our main new result is the existence of a class of oscillating
models in which the scalar field
is initially slowly rolling, but oscillates at late
times; we show that such models can be consistent with the
observations, but this result is very sensitive to the phase
of the scalar field oscillation at the present.

\section{Evolution of oscillating scalar fields}

\subsection{Previous work}

Consider a minimally-coupled
scalar field, $\phi$, in a potential $V(\phi)$.
The equation of motion for this field in an expanding background is
\begin{equation}
\label{motionq}
\ddot{\phi}+ 3H\dot{\phi} + \frac{dV}{d\phi} =0,
\end{equation}
where the Hubble parameter $H$ is given by
\begin{equation}
\label{H}
H = \left(\frac{\dot{a}}{a}\right) = \sqrt{\rho_T/3}.
\end{equation}
Here $a$ is the scale factor, $\rho_T$ is the total density, and we work in units
for which $8 \pi G = 1$.
The pressure and density of $\phi$
are given by
\begin{equation}
p_\phi = \frac{\dot \phi^2}{2} - V(\phi),
\end{equation}
and
\begin{equation}
\label{rhodense}
\rho_\phi = \frac{\dot \phi^2}{2} + V(\phi),
\end{equation}
respectively, and the equation of state parameter, $w_\phi$,
is defined to be
\begin{equation}
w_\phi = p_\phi/\rho_\phi.
\end{equation}
The density of the scalar field then evolves with the scale factor
as
\begin{equation}
\label{rhoevolve}
\rho_\phi \propto a^{-3(1+w_\phi)}.
\end{equation}

Turner \cite{turner} was the first to consider the behavior of such
a field in the limit where it oscillates on a timescale much
shorter than the Hubble time.  He examined power-law potentials
of the form
\begin{equation}
\label{power}
V(\phi) = k \phi^n,
\end{equation}
and he showed that the value
of $w_\phi$, averaged over the oscillation period, is
\begin{equation}
\label{wturner}
w_\phi = (n-2)/(n+2).
\end{equation}
Thus, the cases $n=2$ and $n=4$ behave like nonrelativistic matter
and radiation, respectively.

Liddle and Scherrer \cite{liddle}
investigated oscillating scalar field models with power-law potentials in the context
of dark energy, noting an interesting set of behaviors:
for a universe dominated by a background fluid with equation of
state parameter $w_B$,
the oscillating solution is an attractor of the equations of motion
only for the case
\begin{equation}
\label{liddlelimit}
n < \frac{2(3+w_B)}{1-w_B},
\end{equation}
while potentials with $n > 2{(3+w_B)}/{(1-w_B)}$ produce a scalar
field trajectory that decreases monotonically to zero without
oscillating.  Liddle and Scherrer provided no
physical explanation for this curious behavior; we give such an
explanation below.

Sahni and Wang \cite{sahni} and Hsu \cite{hsu} examined the possibility that
an oscillating field could give rise to dark energy.  From equation
(\ref{wturner}), it is clear that $w < 0$ requires $n< 2$, and
the production of a dark energy equation of state can only be
achieved if we take
$V = k |\phi|^n$, with $n \ll 1$.  Although potentials
with very small $n$
might appear rather unnatural,
such a potential can arise from a Lagrangian with
a $\phi^2$ potential and
a non-standard kinetic term \cite{hsu}.

Masso et al. \cite{masso}
performed a general investigation of oscillating scalar fields
in arbitrary potentials (as opposed to power laws).  Using
the action-angle formalism, they derived
several very important results, which we will use here.
Consider a scalar field $\phi$ oscillating in an arbitrary
potential $V(\phi)$. The minimum and maximum values of $\phi$ are 
$\phi_1$ and $\phi_2$ respectively.  Then the energy density for this
field must be equal to $\rho_\phi = V(\phi_1) = V(\phi_2)$.
(Note that $\rho_\phi$ changes with time, but it
is assumed to evolve on a timescale longer than the inverse
oscillation frequency).
Masso et al. showed that all of the quantities of interest can
be expressed in terms of the action variable $J$, defined
by
\begin{equation}
\label{J}
J = 2 \int_{\phi_1}^{\phi_2}\sqrt{2(\rho_\phi-V(\phi))}.
\end{equation}
Then the oscillation frequency $\nu$ is
\begin{equation}
\label{nu1}
\nu = \frac{d\rho_\phi}{dJ},
\end{equation}
while the equation of state parameter is
\begin{eqnarray}
1+w_\phi &=& \frac{J}{\rho_\phi}\frac{1}{dJ/d\rho_\phi} ,\\
&=& \frac{2}{\rho_\phi} \frac{\int_{\phi_1}^{\phi_2}[\rho_\phi-V(\phi)]^{1/2} d\phi}
{\int_{\phi_1}^{\phi_2}[\rho_\phi-V(\phi)]^{-1/2} d\phi}.
\end{eqnarray}
(Note that this last result was also derived by Turner \cite{turner}).
These results allow for the investigation of somewhat
more natural (non-power-law) potentials that
might serve as dark energy, and several of these are discussed
in Ref. \cite{masso}.

Finally, Gu \cite{gu} revisited the power-law scenario, and calculated
the conditions necessary for the oscillation frequency to exceed
the expansion rate (our equation \ref{powernu} below).  He used this
condition to place constraints on such power-law models.

\subsection{Evolution of oscillation frequency}

Implicit in all of these discussions is the assumption that
the oscillation frequency is much greater than the Hubble expansion
rate.  We now consider this condition in more detail.  The relevant
quantity is the ratio of oscillation frequency to Hubble parameter,
$\nu/H$.  If $\nu/H \gg 1$, then the parameters of interest
can be averaged over the oscillation period, as in Refs.
\cite{liddle,turner,sahni,
hsu,masso,gu}.  On the other hand, when $\nu/H \ll 1$, the field effectively
ceases to oscillate, and an alternative treatment is required.

Consider the evolution of $\nu/H$.  From equations (\ref{J})
and (\ref{nu1}) we can write
\begin{equation}
\nu = \left[\int_{\phi_1}^{\phi_2}\sqrt{\frac{2}{\rho_\phi - V(\phi)}}d\phi\right]^{-1}.
\end{equation}
This equation, along with the expression for $H$ given by
equation (\ref{H}), gives us
\begin{equation}
\label{nuH}
\nu/H = \left[\int_{\phi_1}^{\phi_2}\sqrt{\frac{2\rho_T}{3(\rho_\phi - V(\phi))}}d\phi\right]^{-1}.
\end{equation}
For the case of the power-law potential, $V(\phi) = k \phi^n$, the expression for
$\nu$ can be integrated exactly \cite{masso}, giving (see also \cite{gu})
\begin{equation}
\label{powernu}
\nu/H = \frac{\sqrt{3}n\Gamma(1/2 + 1/n)}{2 \sqrt{2\pi} \Gamma(1/n)}\sqrt{\frac{\rho_\phi}{\rho_T}}
\frac{1}{\phi_{max}},
\end{equation}
where $\phi_{max} = \phi_1 = \phi_2$, since the potential is symmetric.
For other symmetric potentials, we can use equation (\ref{nuH}) to derive an order
of magnitude estimate:
\begin{equation}
\label{nuapprox}
\nu/H \sim \sqrt{\frac{\rho_\phi}{\rho_T}} \frac{1}{\phi_{max}}.
\end{equation}
This result is clearly consistent with the exact result for power-law potentials
given by equation (\ref{powernu}).

Consider first the case where the scalar field itself dominates
the expansion, so that $\rho_T \approx \rho_\phi$.  In this
case, we have $\nu/H \sim 1/\phi_{max}$.  Since $\rho_\phi$ always
decreases with time, $\phi_{max}$ is also a decreasing function
of time, so that $\nu/H$ always increases with time.

The other possibility is that the universe is dominated by
a background fluid (e.g., matter or radiation) with density
$\rho_B$.  In this case, we have $\nu/H \sim \sqrt{\rho_\phi/\rho_B}
(1/\phi_{max})$.  We can again assume that $\phi_{max}$ decreases
with time.  In order for $\phi$ to serve as a plausible dark
energy candidate, its density must decrease more slowly with
the expansion than the density of matter or radiation.  Hence,
we can conclude that $\rho_\phi/\rho_B$ increases with
time, and once again we have that $\nu/H$ increases with time.

Thus, for any model in which
$\phi$ serves as the dark energy today, $\nu/H$ is an increasing function of time,
regardless of whether the universe is background-dominated or
$\phi$-dominated.

Before examining the implications of this result,
we reconsider the behavior discussed in Ref. \cite{liddle} and described
by equation (\ref{liddlelimit}).  Equations (\ref{wturner}) and
(\ref{liddlelimit}), taken together, imply that scalar field oscillations
are an attractor in a universe dominated by a background fluid only for the case where
\begin{equation}
\label{wphiliddle}
w_\phi < 1/2 + w_B/2.
\end{equation}
Further, for a power-law potential given by equation (\ref{power}),
in the background-dominated case,
equation (\ref{nuapprox}) can be expressed as
\begin{equation}
\nu/H \sim \rho_B^{-1/2}\rho_\phi^{1/2 - 1/n}.
\end{equation}
Applying equation (\ref{wturner}), this becomes
\begin{equation}
\nu/H \sim \rho_B^{-1/2}\rho_\phi^{w_\phi/(1+w_\phi)},
\end{equation}
which can be rewritten in terms of the scale factor (using equation
(\ref{rhoevolve}) for $\rho_\phi$ and the corresponding equation
for $\rho_B$) as
\begin{equation}
\label{nua}
\nu/H \sim a^{3(1/2 + w_B/2 - w_\phi)}.
\end{equation}
A comparison of equations (\ref{nua}) and (\ref{wphiliddle})
provides a physical explanation for the bound noted by Liddle
and Scherrer \cite{liddle}.  When this bound (equation \ref{liddlelimit})
is not satisfied, equation (\ref{nua}) indicates that
$\nu/H$ is a {\it decreasing} function of time.  Thus,
at late times, the oscillating solution cannot be an attractor.
This is consistent with the behavior noted in Ref. \cite{liddle};
when equation (\ref{liddlelimit}) is not satisfied, the scalar
field decays asymptotically to the minimum of the potential without
oscillating.  These models do not contradict our claim
that in any oscillating scalar field model for dark energy,
$\nu/H$ always increases with time.  The reason is that models
that violate the bound in equation (\ref{liddlelimit}) have
an energy density that decays more rapidly than $\rho_B$; they
are not plausible models for dark energy.

Now consider the possible oscillating scalar field models for dark energy.
The fact that $\nu/H$ increases with time simplifies the classification
of such models.  If we assume that the universe
evolves from a matter-dominated state
to a scalar-field dominated state, there are basically three possibilities,
which depend on the redshift at which $\nu$ becomes greater than $H$:

I.  The oscillation frequency becomes greater than the expansion
rate during the matter-dominated era.  Since $\nu/H$ increases
with time, we always have $\nu/H > 1$ during the quintessence-dominated era,
and we can
use time-averaged quantities to describe the equation of
state of the quintessence.  We will call this the early-oscillation
case.  This is basically the model described
in Refs. \cite{turner,sahni,hsu,masso,gu}, so these models have been exhaustively
explored.

II.  The oscillation frequency is less than $H$
at the present day.  Again, since $\nu/H$ always increases with
time, this implies that $\nu$ is never larger than $H$, so these
models never reach the oscillatory stage.  In the language
of Ref. \cite{CL}, these are thawing models in which
the field has not yet reached the minimum of the potential.
Such models are characterized by a value of $w_\phi$ very close
to $-1$ initially, which increases at late times.  We will
call this the ``no oscillation" case.  Again,
such models have been exhaustively explored (see, e.g., Ref.
\cite{SS} and references therein).

III. The case intermediate between I and II occurs
when $\nu$ becomes greater than $H$ at late times, after
the quintessence density exceeds the matter density.
In this case, the scalar
field displays two distinct
behaviors during the quintessence-dominated era.  At first, when
the oscillation frequency
is less than $H$, the scalar field rolls down the potential and $w_\phi$ increases.
Then,
when the oscillation frequency becomes greater than $H$, the
field oscillates rapidly, and $w_\phi$ takes its time-averaged value
as in Refs. \cite{turner,sahni,hsu,masso,gu}.  This third possibility has
not been previously studied; we will refer to it as the ``late oscillation"
case.

\section{Comparison with Observations}

Although we believe that our classification scheme in the previous section
should describe most possible models for quintessence from oscillating
scalar fields, it is obviously impossible to compare all such models
with the observational data.  We will therefore limit ourselves
to a representative set of models:  the power-law oscillating
scalar field models discussed in Refs. \cite{sahni,hsu,masso,gu}.  As we note
below, we believe that the general features of
the results we derive for these models
will be more generally applicable to other oscillating scalar field models.

Following Ref. \cite{hsu}, we choose the potential to have
the form:
\begin{equation}
V\left( \phi \right)=k\vert\phi\vert^n,
\end{equation}
where $k$ and $n$ are constants.  Then the evolution of the field
depends on three parameters:  $k$, $n$, and the initial value
of $\phi$, which we denote $\phi_i$.
However, the parameter $k$ can be eliminated from
the equation of motion (equation \ref{motionq}) by a suitable rescaling
of $\phi$ and the time $t$.  This rescaling means that we are no longer
using Planck units, but our units of time are irrelevant, since
we always define the ``present day" in terms of the value of $\Omega_\phi$,
rather then $t$.
(Another way to see this is
to note that all of our results depend only on the evolution of $w_\phi$
as a function of $\Omega_\phi$, and $w_\phi(\Omega_\phi)$
depends only on $V^\prime/V$, which
is independent of $k$ \cite{zlatev}).  Thus, the evolution can be
described entirely in terms of $n$ and $\phi_i$.

For a given choice of values of the
two free parameters, $\phi_i$, and  $n$,
we numerically evolve the system from a
starting point in the matter dominated
regime to the point when $\Omega_{m}=0.3$,
which we label as the present time $z=0$.

\begin{figure}
	\epsfig{file=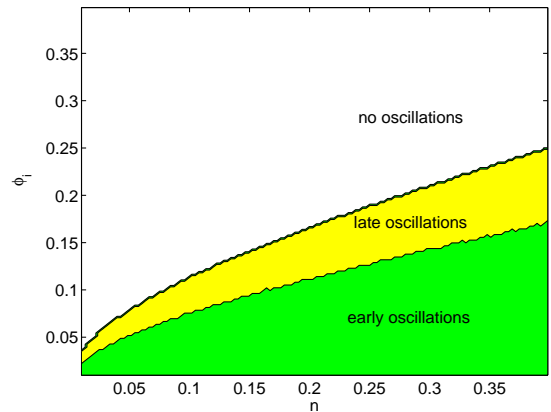,height=55mm}
	\caption
	{\label{oscbehavior_plot}
For the potential $V(\phi) = k|\phi|^n$,
in the parameter space determined by $n$ and the initial magnitude of the scalar field,
$\phi_i$, this plot gives the
regions where the field $\phi$ oscillates prior to dominating the expansion (early
oscillations), after dominating the expansion (late oscillations), or does
not oscillate at all prior to the present (no oscillations). 
  }
\end{figure}

The relation between the evolution of $\phi$ and the two input parameters
is illustrated in Fig. \ref{oscbehavior_plot}.
In our numerical simulations, we define oscillations to commence
when the scalar field crosses $\phi=0$.  This corresponds to $w_\phi$
increasing from $-1$ to $+1$ and marks the point at which $dw_\phi/dz$
first changes sign.  (Note that the oscillation frequency for $w_\phi$
is twice as large as the oscillation frequency for $\phi$).
For early oscillations, this
zero-crossing takes place during the matter-dominated era, while
for late oscillations it takes place during the quintessence-dominated era.
The no oscillation case corresponds to the parameter region in which
zero crossing never occurs up to the present.

\begin{figure}
	\epsfig{file=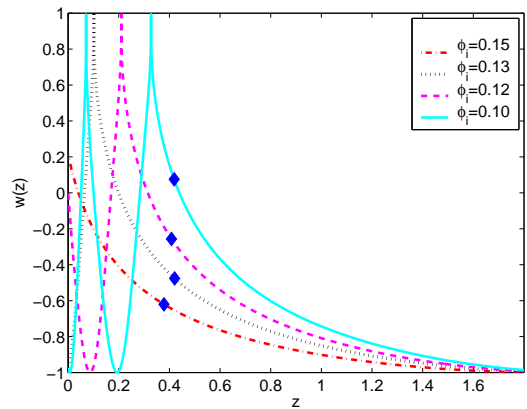,height=55mm}
	\caption{\label{hi_phi}
  Evolution of the quintessence equation of state parameter, $w_\phi$, as a
  function of redshift $z$, in the late oscillation regime.  Curves correspond
  to the indicated values of the initial value of the field, $\phi_i$,
  in the power-law potential $V = k|\phi|^n$ with $n=0.16$.
  Diamonds mark the points at which the density
  of dark energy equals the density of matter.  The curve for $\phi_i = 0.15$
  corresponds to the no oscillation regime and has been added for comparison.}
\end{figure}

For a given $n$, the parameter $\phi_i$ determines when
the oscillations will begin.  From equation (\ref{powernu}) (or
\ref{nuapprox}), it is clear that small $\phi_i$ corresponds
to a larger value of $\nu/H$ for a given ratio of $\rho_\phi$
to $\rho_T$.  Thus, oscillations commence earlier if $\phi_i$
is small and later if $\phi_i$ is large.  For fixed
$n$, sufficiently small
$\phi_i$ will produce the early oscillation case, while sufficiently
large $\phi_i$ gives no oscillations, with late oscillations arising
from intermediate values of $\phi_i$.  This is apparent in Fig.
\ref{oscbehavior_plot}.
Fixing $n$, we see that increasing $\phi_i$ moves the evolution
from early oscillation, through late oscillation, and into no oscillation.
Since the late oscillation case has not been previously
explored, we present a graph, in Fig. \ref{hi_phi}, of
the evolution of $w(z)$ for some of these models.  As claimed,
$w_\phi$ initially evolves slowly upward from $w_\phi=-1$, but then oscillates
rapidly at late times; this transition takes place after the quintessence
field becomes dominant.

\begin{figure}
	\epsfig{file=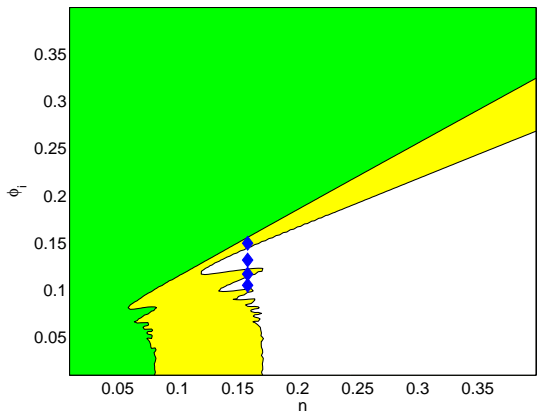,height=55mm}
	\caption
	{\label{likelihood_plot}
	Likelihood plot for the parameters $\phi_i$ and $n$. The $95\%$
	confidence region is green (dark shaded) and the $68\%$ confidence
	region is green+yellow (dark shaded + light shaded). The diamonds
	indicate
	the models displayed in
	Figs \ref{hi_phi}, \ref{mu1_plot}, \ref{mu2_plot}, and \ref{rhophi_plot}.}
\end{figure}

Next, we compare the theoretical luminosity-vs-distance
results to the recent Type Ia Supernovae standard candle
data (ESSENCE+SNLS+HST from \cite{Davis}).
The theoretical distance modulus in our model is calculated in the standard
way:
\begin{equation}
\mu(z) = 5\log_{10}D_L(z) + 42.38 - 5 \log_{10} h,
\end{equation}
where $h$ is the Hubble parameter in units of 100 km sec$^{-1}$ Mpc$^{-1}$,
and $D_L(z)$ is the dimensionless luminosity distance, given by
\begin{equation}
D_L(z) = (1+z) \int_0^z \frac{H_0}{H(z^\prime)} dz^\prime,
\end{equation}
where $H(z^\prime)$, the Hubble parameter at redshift $z^\prime$,
also depends on $\Omega_{m0}$ and the parameters of our quintessence model.
We construct a $\chi^2$ likelihood plot
for $\phi_i$ and $n$ ranging from 0.01 to 0.40 (no further structure emerges from extending the parameter
space).  We take $\Omega_{m0} = 0.3$ and marginalize over the nuisance
parameter $h$.
Fig. \ref{likelihood_plot} shows this
plot.

\begin{figure}
	\epsfig{file=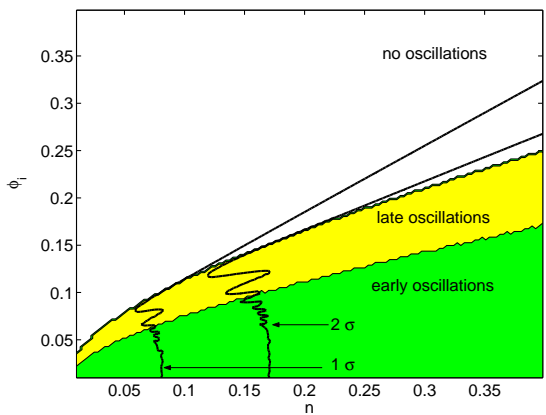,height=55mm}
	\caption
	{\label{likelihood_phi}
The likelihood contours of Fig. \ref{likelihood_plot} superimposed
over the demarcation of the type of oscillatory behavior
from Fig. \ref{oscbehavior_plot}. 
  }
\end{figure}

Some of the allowed regions can be understood in terms of our previous
discussion of the behavior of the field.  In the limit of large $\phi_i$,
the field never oscillates, and it can be effectively treated as
a cosmological constant, which is known to be a good fit to the observational
data.  For sufficiently small $\phi_i$, we are in the ``early oscillation" regime.
In this case, $w_\phi$ takes on a constant value determined by equation
(\ref{wturner}) and independent of $\phi_i$.  It is clear from
equation (\ref{wturner}) that a value
of $w_\phi$ near $-1$ requires a value of $n$ close to zero.
Thus,
we see in Fig. \ref{likelihood_plot} that
for small $\phi_i$, the allowed region is confined to small values
of $n$ (which give $w_\phi$ near $-1$), and the likelihood contours
become vertical lines (indicating that the late-time value of $w_\phi$ is
independent of $\phi_i$ in this limit).

To explore the allowed parameter space further, we
have superimposed the field behaviors outlined in Fig. \ref{oscbehavior_plot}
onto
the likelihood contours displayed in Fig. \ref{likelihood_plot}.
The results are displayed in Fig. \ref{likelihood_phi}.  As noted earlier,
almost the entire ``no oscillation" parameter space is allowed, while
the ``early oscillation" parameter space is confined to the region
for which $w_\phi$ is close to $-1$.  However, we also see that
there is an allowed region in the ``late oscillation" regime; this
is the subset of oscillating models that has not been previously explored.
In these models, the field is initially slowly rolling down the potential,
with increasing $w_\phi$; it then transitions into rapid oscillations
at late times.  An interesting feature in this regime
is the non-smooth structure of the confidence contours.  This is a real
effect.

\begin{figure}
	\epsfig{file=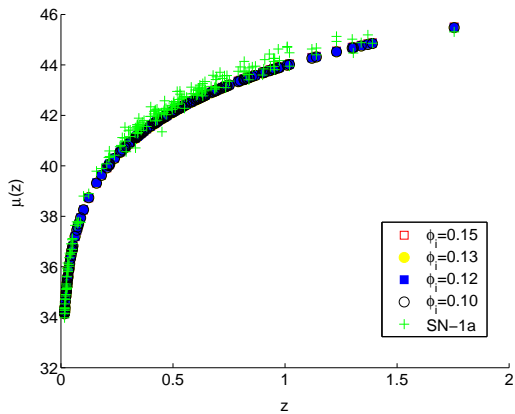,height=55mm}
	\caption
	{\label{mu1_plot}
Distance modulus for the indicated values of $\phi_i$ along with the
supernova values used in our analysis, for $V(\phi) = k|\phi|^n$, with
$n = 0.16$.}
\end{figure}

\begin{figure}
	\epsfig{file=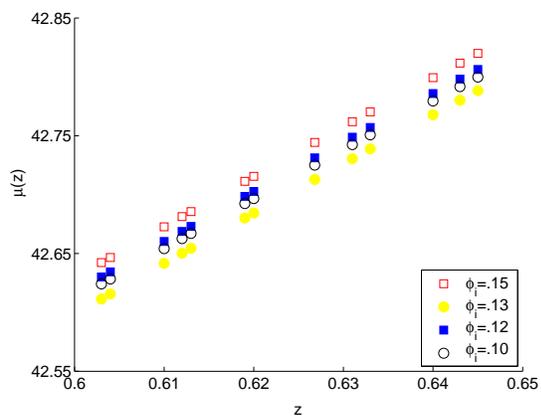,height=55mm}
	\caption
	{\label{mu2_plot}
A blow-up of Fig. \ref{mu1_plot}, illustrating the difference
in behavior of
the four indicated models.}
\end{figure}

\begin{figure}
	\epsfig{file=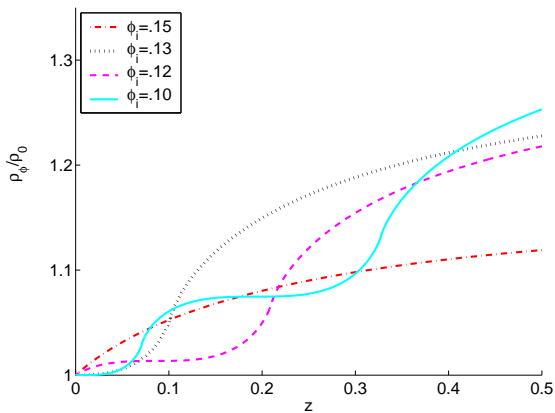,height=55mm}
	\caption
	{\label{rhophi_plot}
Evolution of the scalar field density, normalized to unity
at the present, as a function
of redshift for the four models displayed
in Figs. \ref{hi_phi}, \ref{mu1_plot}, and \ref{mu2_plot}.}
\end{figure}

To illustrate the reason for the non-smooth
confidence contours, we have sampled
four representative models in this regime, denoted by the four diamonds
in Fig. \ref{likelihood_plot}, which correspond to $n=0.16$
and $\phi_i = $0.10, 0.12, 0.13, and 0.15.  (These
are the four models previously displayed in Fig. \ref{hi_phi}).
This illustrates
the complex structure of the allowed region as $\phi_i = 0.10$ and 0.13
are ruled out at 95\% confidence, while $\phi_i = 0.12$ and 0.15 are not.
Next, we display the distance modulus for these four models, along
with the supernova data, in Fig. \ref{mu1_plot}.  Although the individual
models cannot be distinguished here, it is clear that all of
these models lie near the edge of the allowed region because
they tend to produce a distance modulus that is lower than
the observations.  In Fig. \ref{mu2_plot}, we show a blow-up
of Fig. \ref{mu1_plot}.  The smallest values of the distance
modulus are produced for $\phi_i = 0.10$ and 0.13; these are the two
models that are ruled out at 95\% confidence.  The important point
here is that the distance modulus is not a monotonic function
of $\phi_i$.  The reason for this is illustrated in Fig. \ref{rhophi_plot},
where we give the redshift dependence of the
density of the scalar field (which determines
$H(z)$, and therefore, $\mu(z)$).  Comparing
this figure with Fig. \ref{hi_phi}, we see that $H(z)$ depends
crucially on the oscillation phase at the present day.  The two
models for which the field is at a maximum of the potential at present
(thus giving $w_\phi=-1$ at present)
are $\phi_i = 0.10$ and $\phi_i = 0.13$.  These produce a larger
density at moderate redshift and therefore a
smaller integrated
value of $1/H(z)$.  Thus, the complex behavior of our contours
in Figs. \ref{likelihood_plot} and \ref{likelihood_phi} is a
phase effect:  the phase of $\phi$ at present is an oscillating
function of $\phi_i$, and this phase determines whether or not
a given model is excluded.  For early oscillation models the phase
becomes irrelevant, since these models oscillate many times before
the present, and the equation of state parameter takes
on its period-averaged value.


\section{Conclusions}

Our results indicate that oscillating quintessence models, even
for very simple potentials, can display more complex behavior than
has previously been considered.  Previous discussions have centered
on the early oscillation case, in which the oscillation frequency
is always much larger than the expansion rate when the scalar
field dominates the expansion.  Our results show
that the late oscillation case, in which the scalar field transitions
from a slowly rolling regime at early times into oscillations at late
times, can also be consistent with current observations, although
the behavior of these latter models depends crucially on the oscillation
phase of the scalar field at present.  Johnson and Kamionkowksi
\cite{JK} have argued recently that rapidly-oscillating scalar fields
with a time-averaged value of $w_\phi$ less than $0$ are unstable
to the growth of inhomogeneities (see also the argument in
Ref. \cite{Kasuya}).  In this case, only the
late oscillation models would remain as acceptable
oscillating scalar field models for dark energy.

How general are our results?  The conclusion that $\nu/H$
increases with time depends only on the assumption that the scalar
field energy density decays more slowly than the background energy
density, so it should be
generally applicable to any plausible model for dark energy.
On the other hand,
our comparison of models with observations
was restricted to the case of power-law potentials.  However, we would
expect qualitatively similar results for other bound potentials with
a single minimum at $V=0$.  If the minimum of the potential were
at $V_0 > 0$, the analysis is almost identical, except that the
dark energy now has an additional component with constant density
$V_0$.  Such a model would have a values of $w(z)$ closer to $-1$ than
the corresponding model with a minimum at $V=0$, and thus, would
tend to agree more closely with current observations.  One type
of potential that could display completely different
behavior from those studied here is a $V(\phi)$ with multiple minima,
such as the Mexican hat potential discussed in Ref. \cite{masso}.
In this case, the field evolution can be considerably more complex,
as the field can initially oscillate over the entire range of
the potential, but will eventually become trapped in a local
minimum at late times.  In this case the simple analysis presented
here will not apply.

\acknowledgments

R.J.S. was supported in part by the Department of Energy (DE-FG05-85ER40226).

{}


\begin{thebibliography}{99}

\bibitem{Knop}
R.A. Knop, et al., Ap.J. {\bf 598}, 102 (2003).

\bibitem{Riess1}
A.G. Riess, et al., Ap.J. {\bf 607}, 665 (2004).

\bibitem{Wood-Vasey}
W.M. Wood-Vasey, et al., \apj {\bf 666}, 694 (2007).

\bibitem{Davis}
T.M. Davis, et al., \apj {\bf 666}, 716 (2007).

\bibitem{Copeland}
E.J. Copeland, M. Sami, and S. Tsujikawa, Int. J. Mod. Phys. D
{\bf 15}, 1753 (2006).

\bibitem{ratra}
B. Ratra and P.J.E. Peebles,
\prd {\bf 37}, 3406 (1988).

\bibitem{tw}
M.S. Turner and M. White,
\prd {\bf 56}, 4439 (1997).

\bibitem{caldwelletal}
R.R. Caldwell, R. Dave, and P.J. Steinhardt,
\prl {\bf 80}, 1582 (1998).

\bibitem{liddle}
A.R. Liddle and R.J. Scherrer,
\prd {\bf 59}, 023509 (1998).

\bibitem{zlatev}
P.J. Steinhardt, L. Wang, and I. Zlatev,
\prd {\bf 59}, 123504 (1999).

\bibitem{turner}
M.S. Turner, \prd {\bf 28}, 1243 (1983).

\bibitem{sahni}
V. Sahni and L. Wang, \prd {\bf 62}, 103517 (2000).

\bibitem{hsu}
S.D.H. Hsu, Phys. Lett. B {\bf 567}, 9 (2003).

\bibitem{masso}
E. Masso, F. Rota, and G. Zsembinszki,
\prd {\bf 72}, 084007 (2005). 

\bibitem{gu}
J.-A. Gu, arXiv:0711.3606.

\bibitem{KHS}
A. Kurek, O. Hrycyna, and M. Szydlowski, Phys. Lett. B {\bf 659},
14 (2008).

\bibitem{DM}
T. Damour and V.F. Mukhanov, \prl {\bf 80}, 3440 (1998).

\bibitem{dodosc}
S. Dodelson, M. Kaplinghat, E. Stewart, \prl {\bf 85},
5276 (2000).

\bibitem{barosc1}
G. Barenboim, O. Mena, and C. Quigg, \prd
{\bf 71}, 063533 (2005).

\bibitem{barosc2}
G. Barenboim and J. Lykken, Phys. Lett. B {\bf 633},
453 (2006).

\bibitem{fengosc}
B. Feng, M. Li, Y.-S. Piao, and X. Zhang,
Phys. Lett. B {\bf 634}, 101 (2006).

\bibitem{linderosc}
E.V. Linder, Astropart. Phys. {\bf 25}, 167 (2006).

\bibitem{CL}
R.R. Caldwell and E.V. Linder, \prl {\bf 95}, 141301 (2005).

\bibitem{SS}
R.J. Scherrer and A.A. Sen, \prd {\bf 77}, 083515 (2008).

\bibitem{JK}
M.C. Johnson and M. Kamionkowski, arXiv:0805.1748.

\bibitem{Kasuya}
S. Kasuya, Phys. Lett. B {\bf 515}, 121 (2001).

\end{thebibliography}
\end{document}